# Control of Power in Parity-Time -Symmetric Lattices

Maksim Kozlov[1] and G. P. Tsironis[2,3,4]

[1]*Center for Energy Research, National Laboratory Astana, Nazarbayev University, Astana, 010000, Republic of Kazakhstan*

[2]*Department of Physics, School of Science and Technology, Nazarbayev University, Astana, 010000, Republic of Kazakhstan*

[3]*Department of Physics, University of Crete, P. O. Box 2208, Heraklion, 71003, Greece*

[4]*Institute of Electronic Structure and Laser, Foundation for Research and Technology-Hellas, P. O. Box 1527, Heraklion, 71110, Greece*

maksim.kozlov@nu.edu.kz

We investigate wave transport properties of Parity-Time (PT) symmetric lattices that are periodically modulated along the direction of propagation. We demonstrate that in the regime of unbroken PT-symmetry the system Floquet-Bloch modes may interfere constructively leading to either controlled oscillations or linear power absorption and amplification occurring exactly at the phase transition point. The differential power response is effected by the overlap of the gain and loss system distribution with wave intensity pattern that is formed through Rabi oscillations engaging the coupled Floquet-Bloch modes.

A special class of non-Hermitian Hamiltonians that are invariant under simultaneous Parity and Time reversal and possess purely real eigenvalue spectrum was initially investigated in the context of quantum mechanics [1]. The PT-symmetry condition requires the real part of the complex potential in the Schrödinger equation to be symmetric while its imaginary part representing the gain-loss distribution to be antisymmetric. These systems undergo a PT-symmetry breaking transition at some critical peak value of the imaginary part of the potential. Above this threshold value the spectrum ceases to be real, implying instability for the quantum system. The idea of PT-symmetry has been applied to many areas such as cavity quantum electrodynamics [2], classical mechanics [3], magnetohydrodynamics [4, 5] yet perhaps the most intriguing features of PT-symmetry are being investigated in the context of optics [6-17]. Optical systems with PT-invariant complex refractive index supporting waves propagating with constant power were predicted theoretically [6, 7] and demonstrated experimentally [8, 9]. The features of the linear regime include double refraction, power oscillations, nonreciprocal diffraction [6-9], amplification of Goos-Hänchen effect [10] and unidirectional invisibility [11-13]. On the other hand implementation of unidirectional dynamics [14, 15], nonreciprocal soliton scattering [16] and switching [17] require combined action of PT-symmetry and nonlinearity. Experimental studies of active PT-symmetric electric circuits [18] and theoretical investigations of nonlinear magnetic metamaterials [19] were also reported recently.

A special class of configurations with potential applications in optics involves longitudinally modulated PT-symmetric structures [20-24]. Among other effects, optical Rabi oscillations, i.e. resonant power transitions between different light modes originally proposed [25, 26] and observed [27] in Hermitian waveguides and wave-guide arrays were also predicted in periodically modulated in the direction of propagation PT-symmetric waveguides [28]. In the present work we focus exactly on these phenomena, viz. on propagation dynamics in complex PT-symmetric wave-guide arrays periodically modulated along direction of propagation. We show that in the regime of unbroken PT-symmetry when all Floquet-Bloch (FB) modes remain stable, mode interference results in a wealth of behaviors including damping, amplification or even unlimited amplification of incident beam power distributed linearly in the propagation direction. The different power regimes depend on the resonant frequency overlap of the gain-loss distribution with the intensity pattern formed through the interference of Rabi-coupled FB modes. A noteworthy feature is the appearance of an intensity locked region at the phase transition point that is bounded by two beams formed through the double refraction of the unique incident beam.

Paraxial beam propagation in the wave-guide arrays periodically modulated along direction of propagation can be described through the Schrödinger equation:

$$i\frac{\partial \psi}{\partial z}+\frac{\partial^2 \psi}{\partial x^2}+U\left[1+\varepsilon\cos(\beta_0 z)\right]\psi = 0, \qquad (1)$$

where $\psi$ is complex amplitude of the beam, $x$ and $z$ are transverse and longitudinal coordinates respectively, the amplitude of longitudinal modulations $\varepsilon$ is taken to be small, i.e. $\varepsilon \ll 1$ and the complex PT-symmetric potential is given by $U = V\cos(2\pi x/D)+iW\sin(2\pi x/D)$. The real part of $U$ that describes the refractive index profile is symmetric in the transverse direction while the imaginary part representing the gain-loss distribution is antisymmetric, i.e. fulfills the PT-symmetry condition. The total field distribution in the wave-guide arrays can be approximated by a superposition of normalized FB modes, i.e.

$$\psi(x,z) = \sum_{n=1}^{\infty}\int_{-\pi/D}^{\pi/D} A_n(k,z)\varphi_n(k,x)\exp\left[i\beta_n(k)z\right]dk, \qquad (2)$$

where $\beta_n(k)$ are the propagation constants and the corresponding FB modes are

$$\varphi_n(k,x) = \exp(ikx)\sum_{l=-\infty}^{\infty} u_l^n(k)\exp\left(il\frac{2\pi}{D}x\right), \quad (3)$$

where $n$ is band number and $k$ is Bloch momentum. The mode population coefficients $A_n(k,z)$ are $z$-dependent and can be evaluated through [6, 7]

$$A_n(k,z) = \frac{D}{2\pi}\exp\left[-i\beta_n(k)z\right]\int_{-\infty}^{+\infty}\varphi_n^*(-k,-x)\psi(x,z)dx. \quad (4)$$

The later expression was obtained utilizing the bi-orthogonality condition [6, 7]

$$\int_{-\infty}^{+\infty}\varphi_n^*(-k,-x)\varphi_m(k',x)dx = \frac{2\pi}{D}\delta_{nm}\delta(k-k'). \quad (5)$$

Eqs. (4-5) were derived with additional condition imposed on eigenvectors [29]:

$$u_{l=0}^n(k) > 0, \quad \forall \ n, \ k. \quad (6)$$

For the regime of unbroken PT-symmetry ($W/V$=0.8) the first three bands of the transmission spectra are depicted in Fig. 1(a). As in the case of real periodic potential [26], longitudinal modulation can support Rabi conversions between two different FB modes provided the resonant condition is satisfied; i.e. the frequency of longitudinal modulations $\beta_0$ must be matched to the difference of propagation constants of the corresponding FB modes. The transitions between 1st and 2nd modes for the right and left incidences are schematically shown by green arrows in Fig. 1 (a). Following the standard procedure [26] one can substitute expansion (2) into Eq. (1), apply bi-orthogonality condition (5) and after neglecting off-resonant terms obtain coupled mode equations for evolution of mode population coefficients

$$i\frac{\partial A_n(k,z)}{\partial z} = \frac{\varepsilon}{2}A_m(k,z)M_{nm}\exp\left[-i(\beta_n - \beta_m - \beta_0)z\right],$$
$$i\frac{\partial A_m(k,z)}{\partial z} = \frac{\varepsilon}{2}A_n(k,z)M_{mn}\exp\left[i(\beta_n - \beta_m - \beta_0)z\right]$$

$$(7)$$

where the coupling coefficients given by the overlap integral

$$M_{nm}(k) = -\int_{-D/2}^{D/2}\varphi_n^*(-k,-x)U\varphi_m(k,x)dx \quad (8)$$

remain purely real and positive if condition (6) is satisfied. Coupling coefficients for 1→2 transitions in the under-critical ($V$=1 and $W$=0.8) regime are shown in Fig. 1(b). The striking difference with the case of Hermitian (real) Hamiltonians is asymmetry of the coupling coefficients with respect to $k$ $M_{nm}(k) \neq M_{nm}(-k)$ and index interchange $M_{nm}(k) \neq M_{mn}(k)$.

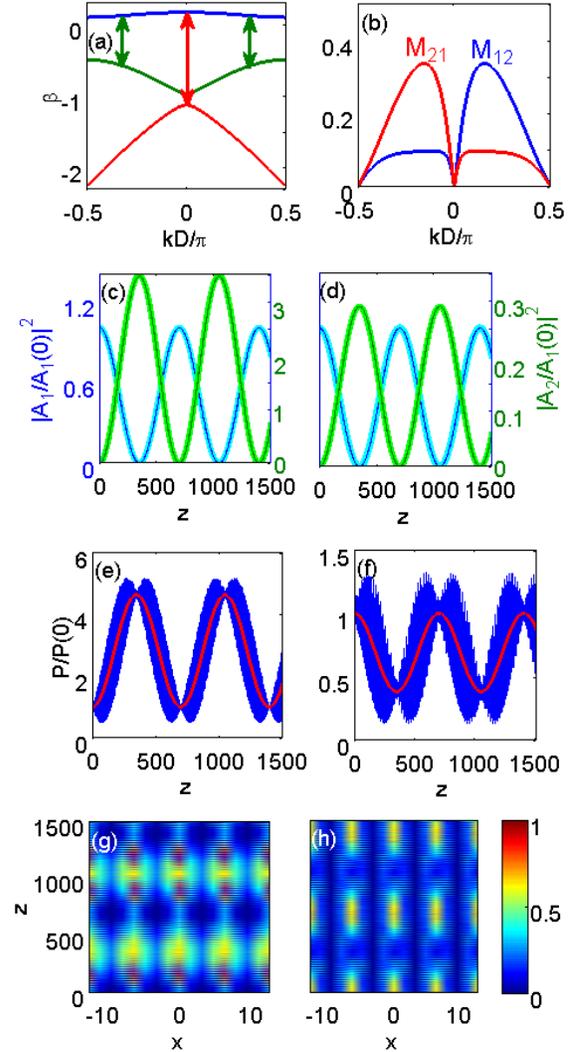

**Fig. 1.** (a) The first 3 bands of the transmission spectra for unbroken PT-symmetry ($W/V$=0.8); green arrows denote 1→2 transitions for tilted incidence and red arrow denote 1→3 transition for normal incidence. (b) – (h) Dynamics of 1→2 transition for $V$=1, $W$=0.8, $\varepsilon$=0.05; (b) coupling coefficients $M_{12}$ (blue curve) and $M_{21}$ (red curve); (c) and (d) evolutions of the 1st (blue curves) and 2nd (green curves) mode populations for left and right tilted incidences respectively; light-colored thick lines - analytical results and dark-colored thin lines - numerical results; (e) and (f) evolution of power normalized to input power for left and right tilted incidences respectively; red curves - analytical results, blue curves - numerical results; (g) and (h) numerically found intensity evolution patterns for the left and right tilted incidences respectively.

However the coupling coefficients remain invariant under simultaneous index interchange and $k$ reversal $M_{nm}(k) = M_{mn}(-k)$ resulting from their definition (8) and the fact that they are real. In Figs. 1(c) and (d) we depict the evolution of the 1st (blue curves) and 2nd (green curves) mode populations for the case when the resonance condition $\beta_1 - \beta_2 - \beta_0 = 0$ is satisfied and only the 1st mode $\varphi_1(k_0, x)$ is excited at the input. Hereinafter the amplitude of longitudinal modulations is taken to be $\varepsilon = 0.05$. Fig. 1(c) and (d) correspond to the left ($k_0D/2\pi = -0.15$) and right ($k_0D/2\pi = 0.15$) tilted incidence respectively. The light-colored thick lines represent analytical results evaluated through solution of Eq. (7) whereas the dark-colored thin lines were obtained from Eq. (4) with field distribution $\psi(x,z)$ found by direct numerical solution of Eq. (1). The mode populations oscillate with frequency $\beta_{Rb} = \varepsilon\sqrt{M_{21}M_{12}}/2$ which is identical to the frequency of Rabi oscillations in real periodic potential [26]. However the asymmetry of coupling coefficients results in unequal amplitudes of mode population oscillations which ratio is given by $\max(|A_2|^2)/\max(|A_1|^2) = M_{21}/M_{12}$.

The physical reason of different amplitudes is that the exchange of energy between two FB modes is augmented by their amplification (damping) due to the positive (negative) overlap of intensity pattern with the gain-loss distribution. For example in the case of the left tilted incidence ($M_{21} > M_{12}$) in the first half-period 2$^{nd}$ mode accepts power from the 1$^{st}$ mode and is also amplified by the positive overlap whereas in the second half-period it loses its power to the 1$^{st}$ mode yet it is also damped by negative overlap. The 1$^{st}$ mode in the first half period loses its power to the 2$^{nd}$ mode yet it is also amplified by the positive overlap whereas in the second half-period it accepts power from the 2$^{nd}$ mode but is also damped by the negative overlap.

The evolution of the total power is given by

$$P(z)/P(0) = 1 + \eta_{21}(k_0)\sin^2(\beta_{Rb}z), \qquad (9)$$

where amplitude of power variations is given by $\eta_{21}(k) = \theta_{21}M_{21}/M_{12} - 1$ and

$$\theta_{21} = \int_{-D/2}^{D/2}|\varphi_2(k,x)|^2 dx \bigg/ \int_{-D/2}^{D/2}|\varphi_1(k,x)|^2 dx \quad \text{represents the}$$

ratio of power contents of different modes. In contract with the case of real potential this ratio is not necessarily equal to one because FB modes were normalized using bi-orthogonality condition (5). The power evolution given by Eq. (9) is depicted by red curves in Fig. 1 (e) and (f) that correspond to the left ($k_0D/2\pi = -0.15$) and right ($k_0D/2\pi = 0.15$) tilted incidence respectively. For the left tilted incidence $\eta_{21} > 0$. In this case both modes and total power are amplified during the first half period and dumped in the second half period. For the right tilted incidence $\eta_{21} < 0$ so dynamics is opposite: both modes and total power are damped during the first half period and amplified in the second half period. The blue curves in Fig. 1 (e) and (f) represent power evolution obtained by direct numerical solution of Eq. (1). Numerically obtained power exhibits fast oscillations around analytical solution represented by Eq. (9) due to the non-orthogonality of FB modes [6, 7]. These fast oscillations diminish in the vicinity of points where population of one of the modes becomes small. The dynamics of the 1→2 transitions that is described above can be also observed in Figs. 1 (g) and (h) depicting numerically found intensity evolution patterns for the cases of the left and right tilted incidences respectively. In the case of normal incidence $M_{21} = 0$ and $M_{12} = 0$ therefore the system cannot support 1→2 transitions.

The coupling coefficients for 1→3 transitions are shown in Fig. 2(a). The dynamics in the case of 1→3 transitions is very similar for the tilted incidences and therefore is not shown here. The only difference with respect to the 1→2 transitions is that $\theta_{31}M_{31}/M_{13} > 1$ for the right tilted incidence and $\theta_{31}M_{31}/M_{13} < 1$ for the left tilted incidence. Therefore the power is initially increasing for the right tilted incidence and vice versa. However for the normal incidence ($k = 0$) the dynamics of 1→3 transitions is different. The transition between 1$^{st}$ and 3$^{nd}$ modes for the normal incidence is schematically shown by red arrow in Fig. 1 (a). Coefficients $M_{31}(k=0) \neq 0$ and $M_{13}(k=0) \neq 0$ therefore 1→3 transitions are supported. The evolution of the 1$^{st}$ (blue curves) and 3$^{rd}$ (green curves) mode populations for the case of normal incidence are depicted in Fig. 2(b).

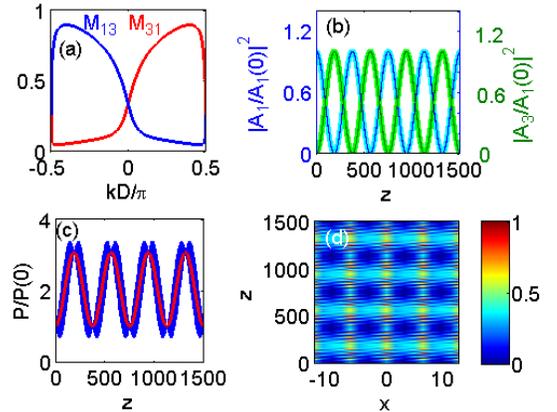

**Fig. 2.** Dynamics of 1→3 transition for $V=1$, $W=0.8$, $\varepsilon=0.05$; (a) coupling coefficients $M_{13}$ (blue curve) and $M_{31}$ (red curve); (b) evolutions of the 1$^{st}$ (blue curves) and 3$^{rd}$ (green curves) mode populations for the normal incidence; light-colored thick lines - analytical results, dark-colored thin lines - numerical results; (c) evolution of power normalized to input power for the normal incidence; red curve - analytical result, blue curve - numerical result; (d) numerically found intensity evolution pattern for the normal incidence.

The resonance condition $\beta_1 - \beta_3 - \beta_0 = 0$ was satisfied and only the 1$^{st}$ mode $\varphi_1(k=0,x)$ was excited at the input. The light-colored thick lines represent analytical results and dark-colored thin lines were obtained by substituting numerically found field distribution $\psi(x,z)$ into Eq. (4). The amplitudes of mode oscillations are equal in this case since $M_{31}(k=0)/M_{13}(k=0) = 1$. Evolution of power for 1→3 transition governed by the Eq. (9) with index 2 replaced by 3 is shown in Fig. 2 (c) by the red curve. Power oscillations result from unequal power content of different modes, viz. $\theta_{31} \neq 1$. The blue curve in Fig. 2 (c) represents power evolution obtained by direct numerical solution of Eq. (1). The fast oscillations of the numerical solution result from non-orthogonality of FB modes. The dynamics of the $1 \rightarrow 3$ transitions that is described above can be also observed in Fig. 2 (d) depicting numerically found intensity evolution pattern for the case of normal incidence.

The most intriguing feature involves the propagation in the phase transition regime when $W/V=1$. The first three bands of the transmission spectra in this regime ($V=W=1$) are depicted in Fig. 3(a) and the coupling coefficients for $1 \rightarrow 2$ transitions are shown in Fig. 3 (b). One can observe that $M_{12} = 0 \ \forall \ k \leq 0$. It follows then from Eqs. (7) that for the left tilted incidence $A_1 = const$, while

$$A_2(k,z) = \frac{\varepsilon}{2} A_1(k) M_{21}(k) \frac{1 - \exp[i(\beta_1 - \beta_2 - \beta_0)z]}{\beta_1 - \beta_2 - \beta_0} + A_2(k,0)$$

(10)

Substituting the above expression for $A_2$ and Eq. (3) for $\phi_2$ into the $n=2$ term of Eq. (2) and integrating over $k$ by the method of residues one finds that

$$\int_{-\pi/D}^{\pi/D} A_2(k,z) \varphi_2(k,x) \exp[i\beta_2(k)z] dk = \frac{\pi \varepsilon}{2\chi} A_1(k_0) M_{12}(k_0) \times rect\left(x, -z \frac{\partial \beta_1}{\partial k}\bigg|_{k=k_0}, -z \frac{\partial \beta_2}{\partial k}\bigg|_{k=k_0}\right)$$

$$\times \exp[i\beta_2(k_0)z] \varphi_2(k_0,x) + \int_{-\pi/D}^{\pi/D} A_2(k,0) \varphi_2(k,x) \exp[i\beta_2(k)z] dk$$

(11)

where $rect(x,x_1,x_2) = 1$ if $x \in [x_1, x_2]$ and $rect(x,x_1,x_2) = 0$ if $x \notin [x_1, x_2]$ and $\chi = (\partial \beta_1/\partial k - \partial \beta_2/\partial k)|_{k=k_0}$. In the derivation of Eq. (11) we assumed that the resonance condition $\beta_1(k_0) - \beta_2(k_0) - \beta_0(k_0) = 0$ is satisfied and neglected diffraction terms that are proportional to $(k-k_0)^2$.

The propagation dynamics at the phase transition regime in the case when the resonance condition $\beta_1 - \beta_2 - \beta_0 = 0$ is satisfied and modes with single Bloch wave-number $k_0$ are excited at the input is straightforward: population of the 1st mode remains constant while population of the 2nd mode increases according to expression obtained by taking limit of Eq. (10) at $k \rightarrow k_0$:

$$|A_2(k_0,z)| = \left|\frac{i\varepsilon}{2} A_1(k_0) M_{21}(k_0) z - A_2(k_0,0)\right|$$

(12)

The 2nd mode is amplified at constant rate due to the constant population of the 1st mode and positive overlap of intensity pattern with gain-loss distribution. If $A_2(k_0,0) = 0$ the population of the 2nd mode growth linearly and its power $\sim |A_2(k_0,z)|^2$ growths quadratically.

In the following we will focus on the evolution of the Gaussian incident beam which can be represented as superposition of modes with different Bloch wave-numbers. The propagation dynamics of the left tilted incident beam $\psi(x,z=0) = \exp(-x^2/\sigma^2)\exp(ik_0 x)$ ($\sigma = 60$, $k_0 D/2\pi = -0.15$), is shown in Fig. 3 (c)-(f). The evolution of the 1st (blue curves) and 2nd (green curves) mode populations at $k = k_0$ are depicted in Fig. 3(c). The light-colored thick lines represent analytical results given by Eq. (12) and dark-colored thin lines were obtained by substituting numerically found field distribution $\psi(x,z)$ into Eq. (4). The population of the 1st mode remains constant upon propagation and the population of the 2nd mode growths almost linearly [small deviation of linear growth at the beginning is due to the $A_2(k_0,0) \neq 0$ term of Eq. (12)]. The numerically obtained evolution of intensity pattern is shown in Fig. 3 (d). We observe the phenomenon of double refraction which manifests itself as splitting of the single incident beam into two diverging beams propagating with the group velocities of the 1st and 2nd modes, i.e. $\partial \beta_1/\partial k (k=k_0)$ and $\partial \beta_2/\partial k (k=k_0)$ respectively [6, 7]. In addition to the double refraction the intensity pattern exhibits an intensity-locked region which develops due to the positive overlap of intensity pattern with gain-loss distribution. Intensity oscillations in this region conform to the periodicity of the lattice while the amplitude of oscillations remains constant between the two diverging beams. The amplitude of the intensity oscillation in this region remains constant upon propagation whereas its width growth linearly as distance between diverging beams increases. This intensity locked region corresponds to the first term on the RHS of Eq. (11). Its intensity (absolute value squared) at the output plane (z=400) is shown in Fig. 3

(e) by the red curve along with numerically found intensity (blue curve). Its power which was obtained by integration of intensity over x

$$P(z)/P_1(k_0) = \varsigma_{21}(k_0)\varepsilon^2 z/D \quad (13)$$

$[\varsigma_{21}(k) = \theta_{21}(\pi M_{21})^2 (\partial\beta_1/\partial k - \partial\beta_2/\partial k)/4$ is coefficient of linear growth and $P_1(k) = |A_1(k,z=0)|^2 \int_{-D/2}^{D/2}|\varphi_1(k,x)|^2 dx$ is initial power of the 1st mode per one cell] is shown by red curve in Fig. 3 (f).

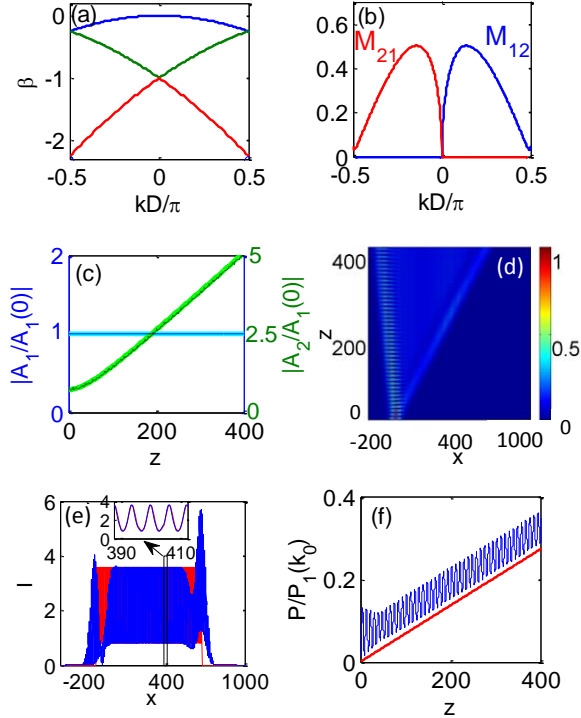

**Fig. 3.** (a) The first 3 bands of the transmission spectra at phase transition point ($V=W=1$); (b) – (f) Dynamics of $1\rightarrow 2$ transition $\varepsilon=0.05$, $\psi(x,z=0)=\exp(-x^2/\sigma^2)\exp(ik_0 x)$, $k_0 D/2\sigma=-0.15$, $\sigma=60$, (b) coupling coefficients $M_{12}$ (blue curve) and $M_{21}$ (red curve); (c) evolutions of the 1st (blue curves) and 2nd (green curves) mode populations; light-colored thick lines - analytical results, dark-colored thin lines - numerical results; (d) numerically found intensity evolution patterns; (e) numerical (blue curve) and analytical (red curve) output (z=400) intensity profile; inset shows zoom area; (f) evolution of power normalized to initial power of the 1st mode per one cell at $k=k_0$; red curves - analytical results, blue curves - numerical results.

The blue curve in Fig. 3 (f) represents numerically found total power of the beam. As previously the fast oscillations result from mode beating while linear amplification is due to the linearly increasing width of intensity locked region.

In the case of right tilted incidence the linear amplification can be implemented through $1\rightarrow 3$ transition since $M_{21}=0 \ \forall \ k\geq 0$, yet $M_{31}>0 \ \forall \ k>0$ and $M_{13}=0 \ \forall \ k\geq 0$. Evolution of 3rd mode population, power and field for $1\rightarrow 3$ transition is governed by the Eqs. (10-13) with index 2 replaced by 3. In the case of normal incidence (k=0) 2nd and 3rd bands coalesce forming an exceptional point. Coresponding FB states become identical and self-orthogonal at this point [30]. This implies that in order for these states to remain normalized to unity their amplitudes (components of eigenvectors $u_l^2$) should become extremely large as $k\rightarrow 0$. Our coupled mode analysis fails at $k=0$ because self-orthogonality precludes normalization of FB modes. However direct simulation of beam propagation shows that the amplitude of the intensity locked region remains finite and continuously changes as $k\rightarrow 0$ because the product of infinitely large $u_l^2$ with diminishing coupling coefficients $[M_{21}(k=0)=0]$ remains finite and continuous.

Importantly, in the regime of unbroken PT-symmetry ($W<V$) one can manipulate both sign and magnitude of power oscillations by changing the angle of incidence and corresponding adjustment of longitudinal frequency modulations to satisfy the phase matching condition. This can be observed in Fig. 4 (a) that depicts amplitudes of power variations $\eta_{21}$ and $\eta_{31}$ vs. Bloch momentum $k$ for $W/V=0.8$. However at the phase transition regime $W=V$ only the rate of always positive linear power amplification can be manipulated through angular dependence of linear growth coefficients $\varsigma_{21}$ or $\varsigma_{31}$. Fig. 4 (b) shows $\varsigma_{21}$ and $\varsigma_{31}$ vs. $k$ for $W=V$.

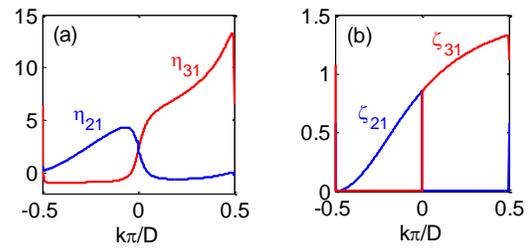

**Fig. 4.** (a) Amplitudes of power variations $\eta_{21}$ and $\eta_{21}$ vs. $k$ for $W/V=0.8$; (b) linear growth coefficients $\varsigma_{21}$ and $\varsigma_{31}$ vs. $k$ for $W=V$.

The phenomena described so far are not particular to 1D lattices. We have also investigated propagation in longitudinally modulated 2D PT-symmetric lattices with potential

$$U = V\left[\cos\left(\frac{2\pi x}{D_x}\right) + \cos\left(\frac{2\pi y}{D_y}\right)\right]$$
$$+ iW\left[\sin\left(\frac{2\pi x}{D_x}\right) + \sin\left(\frac{2\pi y}{D_y}\right)\right],$$

and focused particularly on the phase transition regime ($V=W=1$). The amplitude of the longitudinal modulations is $\varepsilon=0.05$, while normally incident beam profile is given by $\psi(x, y, z=0) = \exp\left[-(x^2 + y^2)/\sigma^2\right]$ and the frequency of longitudinal modulations satisfy resonance condition $\beta_1(k_x, k_y) - \beta_2(k_x, k_y) - \beta_0 = 0$ ($k_x = 0$, $k_y = 0$). Figure 5 (a) depicts the intensity pattern at the output plane ($z=480$). The intensity locked regions in the 2D case correspond to two "ridges" that are directed along x and y axis. Intensity oscillations in this "ridges" conform to the periodicity of the lattice while the amplitude of oscillations remains constant between the beams that diverge in the x and y direction from the main beam propagating along z axis.

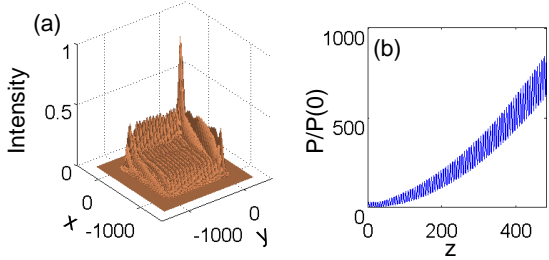

**Fig. 5.** Numerically simulated propagation dynamics of normally incident beam in longitudinally modulated 2D PT-symmetric lattice at phase transition point. Frequency of longitudinal modulations satisfy resonance condition $\beta_1(k_x,k_y)-\beta_2(k_x,k_y)-\beta_0=0$, ($k_x=0,k_y=0$) (a) intensity pattern at output plane ($z=480$); (b) beam power vs. propagation distance.

The flat "plateau" region between two "ridges" develops due to the secondary generation of intensity locked regions in perpendicular directions: x ridge radiates in y direction and y "ridge" radiates in x direction. The development of this square "plateau" region is responsible for quadratic amplification of power shown in Fig. 5 (b).

While power amplification has been predicted in the gain-loss media [31, 32], in the present work we have investigated power control phenomena in the complex PT-symmetric wave-guide arrays that are periodically modulated along direction of propagation. In the regime of unbroken PT-symmetry we demonstrated that at certain resonant frequencies of longitudinal modulation, interference of Rabi-coupled FB modes can result in periodic oscillations as well as linear amplification of the absorbed power due to the overlap of the gain-loss distribution with the wave intensity pattern. Linear power growth that occurs only at the phase transition point is characterized by the development of the intensity locked region between the two diverging beams induced through double refraction. In the latter case only the rate of power amplification can be controlled through the angular dependence of the coupling coefficients. In the under-critical regime both sign and magnitude of periodic power variations can be manipulated through the incidence angle of the beam. In both cases the frequency of longitudinal modulations should be adjusted accordingly to satisfy the phase matching condition. We used coupled mode theory to analyse quantitatively the beam dynamics and verified the results numerically. We found similar effects in the 2D longitudinally modulated PT-symmetric lattices. The phase transition point in the 2D case is characterized by the development of perpendicular intensity-locked "ridges" with a flat "plateau" region between them and quadratic amplification of the power. The efficient power control technique introduced here may have significant applications in PT-symmetric optics as well as in gain-loss metamaterials and metasurfaces.

We acknowledge partial support from the Ministry of Education and Science of the Republic of Kazakhstan under contract 339/ 76−2015 and by the European Union-Seventh Framework Program (FP7-REGPOT-2012-2013-1) under Grant No. 316165.